\documentclass[apj]{emulateapj}
\usepackage{apjfonts}
\usepackage[usenames,dvipsnames]{xcolor}
\usepackage{slantsc}
\usepackage[T1]{fontenc}
\definecolor{blue}{rgb}{0,0,1}
\usepackage{CJK}
\usepackage{multirow}
\usepackage{color}

\newcommand{\tabincell}[2]{\begin{tabular}{@{}#1@{}}#2\end{tabular}}

\newcommand{\bdv}[1]{\mbox{\boldmath$#1$}}
\usepackage[hyperfootnotes=false]{hyperref}
\hypersetup{
  colorlinks,
  citecolor=blue,
  linkcolor=blue,
  urlcolor=blue}

\slugcomment{}
\shortauthors{Zhu et al.}
\begin{document}

\title{Planet Sensitivity from Combined Ground- and Space-based Microlensing Observations}
\begin{CJK*}{UTF8}{gkai}

\author{Wei~Zhu~(祝伟)~$^{1,*}$,
Andrew~Gould~$^1$,
Charles~Beichman~$^2$,
Sebastiano~Calchi~Novati~$^{2,3,4,}$\altaffilmark{22},
Sean~Carey~$^{5}$,
B.~Scott~Gaudi~$^1$,
Calen~B.~Henderson~$^{1,6,}$\altaffilmark{23},
Matthew~Penny~$^{1,}$\altaffilmark{24},
Yossi~Shvartzvald~$^{6,}$\altaffilmark{23},
Jennifer~C.~Yee~$^{7,}$\altaffilmark{24}
\centerline{and}
A.~Udalski~$^{8}$,
R.~Poleski~$^{1,8}$,
J.~Skowron~$^{8}$,
S.~Koz{\l}owski~$^{8}$,
P.~Mr{\'o}z~$^{8}$,
P.~Pietrukowicz~$^{8}$,
G.~Pietrzy{\'n}ski~$^{8}$
M.~K.~Szyma{\'n}ski~$^{8}$,
I.~Soszy{\'n}ski~$^{8}$,
K.~Ulaczyk~$^{8}$,
{\L}.~Wyrzykowski~$^{8}$
\centerline{(The OGLE collaboration)}
F.~Abe$^{9}$,
R.~K.~Barry$^{10}$,
D.~P.~Bennett$^{11}$,
A.~Bhattacharya$^{11}$,
I.~A.~Bond$^{12}$,
M.~Freeman$^{13}$,
A.~Fukui$^{14}$,
Y.~Hirao$^{15}$,
Y.~Itow$^{9}$,
N.~Koshimoto$^{15}$,
H.~Ling$^{12}$,
K.~Masuda$^{9}$,
Y.~Matsubara$^{9}$,
Y.~Muraki$^{9}$,
M.~Nagakane$^{15}$,
K.~Ohnishi$^{16}$,
To.~Saito$^{17}$,
D.~J.~Sullivan$^{18}$,
T.~Sumi$^{15}$, 
D.~Suzuki$^{11}$,
P.~J.~Tristram$^{19}$,
N.~Rattenbury$^{13}$,
Y.~Wakiyama$^{9}$,
A.~Yonehara$^{20}$
\centerline{(The MOA collaboration)}
D.~Maoz$^{21}$,
S.~Kaspi$^{21}$,
M.~Friedmann$^{21}$
\centerline{(The Wise group)} }
\affil{
{$^1$ Department of Astronomy, Ohio State University, 140 W. 18th Ave., Columbus, OH  43210, USA} \\
{$^2$ NASA Exoplanet Science Institute, MS 100-22, California Institute of Technology, Pasadena, CA 91125, USA} \\
{$^3$ Dipartimento di Fisica ``E. R. Caianiello'', Universit\'a di Salerno, Via Giovanni Paolo II, 84084 Fisciano (SA), Italy} \\
{$^4$ Istituto Internazionale per gli Alti Studi Scientifici (IIASS), Via G. Pellegrino 19, 84019 Vietri Sul Mare (SA), Italy} \\
{$^5$ Spitzer Science Center, MS 220-6, California Institute of Technology,Pasadena, CA, USA} \\
{$^6$ NASA Jet Propulsion Laboratory, Pasadena, CA 91109, USA} \\
{$^7$ Harvard-Smithsonian Center for Astrophysics, 60 Garden St., Cambridge, MA 02138, USA} \\
{$^8$ Warsaw University Observatory, Al. Ujazdowskie 4, 00-478 Warszawa, Poland} \\
{$^{9}$Solar-Terrestrial Environment Laboratory, Nagoya University, Nagoya 464-8601, Japan}\\
{$^{10}$Astrophysics Science Division, NASA Goddard Space Flight Center, Greenbelt, MD 20771, USA}\\
{$^{11}$Department of Physics, University of Notre Dame, Notre Dame, IN 46556, USA}\\
{$^{12}$Institute of Information and Mathematical Sciences, Massey University, Private Bag 102-904, North Shore Mail Centre, Auckland, New Zealand}\\
{$^{13}$Department of Physics, University of Auckland, Private Bag 92019, Auckland, New Zealand}\\
{$^{14}$Okayama Astrophysical Observatory, National Astronomical Observatory of Japan, 3037-5 Honjo, Kamo- gata, Asakuchi, Okayama 719-0232, Japan}\\
{$^{15}$Department of Earth and Space Science, Graduate School of Science, Osaka University, Toyonaka, Osaka 560-0043, Japan}\\
{$^{16}$Nagano National College of Technology, Nagano 381-8550, Japan}\\
{$^{17}$Tokyo Metropolitan College of Aeronautics, Tokyo 116-8523, Japan}\\
{$^{18}$School of Chemical and Physical Sciences, Victoria University, Wellington, New Zealand}\\
{$^{19}$Mt. John University Observatory, P.O. Box 56, Lake Tekapo 8770, New Zealand}\\
{$^{20}$Department of Physics, Faculty of Science, Kyoto Sangyo University, 603-8555 Kyoto, Japan}\\
{$^{21}$School of Physics and Astronomy, Tel-Aviv University, Tel-Aviv 69978, Israel}}
\altaffiltext{22}{Sagan Visiting Fellow.}
\altaffiltext{23}{NASA Postdoctoral Program Fellow.}
\altaffiltext{24}{Sagan Fellow.}
\altaffiltext{*}{Email address: weizhu@astronomy.ohio-state.edu}

\submitted{ApJ Accepted}

\begin{abstract}
To move one step forward toward a Galactic distribution of planets, we present the first planet sensitivity analysis for microlensing events with simultaneous observations from space and the ground. 
We present this analysis for two such events, OGLE-2014-BLG-0939 and OGLE-2014-BLG-0124, which both show substantial planet sensitivity even though neither of them reached high magnification. This suggests that an ensemble of low to moderate magnification events can also yield significant planet sensitivity and therefore probability to detect planets.
The implications of our results to the ongoing and future space-based microlensing experiments to measure the Galactic distribution of planets are discussed.
\end{abstract}


\section{Introduction} \label{sec:introduction}

Not relying on the light from the target system, microlensing is in principle sensitive to planets at various line-of-sight distances, suggesting that a sample of microlensing planets will be able to tell us the Galactic distribution of planets. However, a problem of the standard microlensing technique is that the mass $M_{\rm L}$ and distance $D_{\rm L}$ of the lens system, together with the lens-source relative proper motion $\mu_{\rm rel}$, are buried within a single observable, the event timescale $t_{\rm E}$,
\[ t_{\rm E} \equiv \frac{\theta_{\rm E}}{\mu_{\rm rel}},\ \]
with
\[ \theta_{\rm E} \equiv \sqrt{\kappa M_{\rm L} \pi_{\rm rel}};\ \kappa \equiv \frac{4G}{c^2\rm AU} \approx 8.14 \frac{\rm mas}{M_\odot};\ \pi_{\rm rel} \equiv {\rm AU}\left(\frac{1}{D_{\rm L}}-\frac{1}{D_{\rm S}}\right)\ .\]
Although a few methods have been used to resolve this degeneracy, they are either ineffective, in the sense that they can only be applied in very rare cases \citep{Gould:2009,Yee:2009,GouldYee:2013}, or strongly biased toward nearby lenses or long-timescale events \citep{Gould:1992}.
Therefore, previous statistical studies based on microlensing planets had to assume some typical values of the lens system, for example, $D_{\rm L}=4$ kpc and $M_{\rm L}=0.3 M_\odot$ \citep[e.g.,][]{Gould:2010,Clanton:2014}. This is appropriate for complementing the demographics of planets by combining with other detection techniques \citep{Gaudi:2012}, but it has prevented microlensing from demonstrating its unique power: deriving the distribution of planets at various Galactic distances.

This situation has been changing with the emergence of space-based microlensing programs. As has long been realized, combining observations from the ground and at least one satellite that is well separated ($\sim$ AU) from Earth is the only effective way to measure the mass and distance of potentially all microlenses \citep{Refsdal:1966,Gould:1994}, because the measurable parameter from such an experiment, the microlens parallax vector $\bdv{\pi}_{\rm E}$,
with amplitude and direction defined by
\begin{equation}
\pi_{\rm E} \equiv \frac{\pi_{\rm rel}}{\theta_{\rm E}};\ \frac{\bdv{\pi}_{\rm E}}{\pi_{\rm E}} = \frac{\bdv{\mu}_{\rm rel}}{\mu_{\rm rel}}\ ,
\end{equation}
relates to $M_{\rm L}$ and $D_{\rm L}$ by
\begin{equation} \label{eq:ml-dl}
M_{\rm L} = \frac{\theta_{\rm E}}{\kappa \pi_{\rm E}};\quad \frac{1}{D_{\rm L}} = \frac{1}{D_{\rm S}} + \frac{\pi_{\rm E}\theta_{\rm E}}{\rm AU}\ ,
\end{equation}
With observations from Earth and only one satellite, there are four allowed choices of $\bdv{\pi}_{\rm E}$. This four-fold degeneracy arises from the fact that the projected position of the source onto the sky as seen from Earth and the satellite can pass by the lens from the same side or different sides. 
Given that the majority of Galactic microlensing events have well measured source distance $D_{\rm S}$ $(\sim 8$ kpc), Equation(\ref{eq:ml-dl}) suggests that dermining $M_{\rm L}$ and $D_{\rm L}$ requires measurements of both $\pi_{\rm E}$ and $\theta_{\rm E}$. For events with detected planets, $\theta_{\rm E}$ can usually be measured via finite source effects \citep{Yoo:2004}. However, measuring the Galactic distribution of planets requires a determination of the distance distribution of all the lenses being probed, not just those with planets. Or, more precisely, the distance distribution of the planet sensitivity of the events containing these lenses. For single-lens events, only a small fraction can yield definite distance measurements even though space-based parallax can be measured \citep{Zhu:2015b}.
Fortunately, \citet{CalchiNovati:2015} showed that individual distances can, in the great majority of cases, be statistically inferred from a combination of $\bdv{\pi}_{\rm E}$ measurements and kinematic priors derived from a Galactic model.
The 2014 pilot program using \emph{Spitzer} as a microlens parallax satellite has demonstrated the feasibility of such a strategy to derive the Galactic distribution of planets \citep{CalchiNovati:2015}, a key ingredient of which is measuring parallaxes to constrain or measure the lens (host star) distances.

To conduct statistical studies with these microlensing events, we will also need a good understanding of the detection efficiency of planets in each individual event. The estimation of detection efficiency, or planet sensitivity, via microlensing has been studied for a long time, both analytically and computationally. Soon after microlensing by planetary systems was proposed by \citet{MaoPaczynski:1991}, \citet{GouldLoeb:1992} estimated that about 20\% microlensing events would show planetary anomalies if all lenses are solar-like systems. A planet may be detected in a microlensing event if the source \citep[or at least some portion of it,][]{Ingrosso:2009} passes over or near a caustic induced by the planet. Hence, the position of the source at a particular time determines what planets the event is sensitive to at that moment. 
The total planet sensitivity of an event is an integration of the sensitivity over the light curve. The light curve may also be broken into segments each with its own planet sensitivity.
The methodology to compute the planet sensitivity was first proposed and used in \citet{Rhie:2000,GaudiSackett:2000} and \citet{Albrow:2000}.

After the search networks were established, microlensing began giving meaningful constraints on the planet occurrence rate. For example, \citet{Gaudi:2002} put a 33\% upper limit on the occurrence rate of Jupiter-mass planets around Bulge M dwarfs after analyzing 43 intensively monitored events; \citet{Gould:2010} for the first time presented the planet frequency beyond the ``snow'' line, using an ensemble of 13 extremely high-magnification ($A>200$) events. The planet frequency from microlensing has also been studied by \citet{Sumi:2010} and \citet{Cassan:2012}.

The high-magnification events in the \citet{Gould:2010} sample are the most efficient for measuring planet frequency because they are each very sensitive to planets, achieving up to 100\% detection efficiency for the largest mass ratios \citep{GriestSafizadeh:1998}. However, such events are extremely rare and difficult to predict in real-time (which is necessary for both ground-based characterization and scheduling \emph{Spitzer}-type observations). Given these difficulties, it is natural to turn to the much larger sample of low-magnification events to measure the planet frequency. While each one is far less sensitive to planets, their number makes up for this deficiency, allowing for a measurement of the planet frequency as in \citet{Gaudi:2002} or \citet{Cassan:2012}.

The difficulty with low-magnification events is that the arguments used by \citet{Gould:2010} to establish an unbiased sample do not apply. However, \citet{CalchiNovati:2015} suggested that any event can be included in the sample to measure the Galactic distribution of planets as long as they are selected for space-based parallax observations without reference to the presence or absence of planets. This idea has been codified for \emph{Spitzer} and for future space-based microlensing experiments in \citet{Yee:2015b}. In brief, events may be chosen either objectively or subjectively. An event is objectively selected if it meets some pre-determined criteria, in which case, the planet sensitivity from the entire event may be considered. For a subjectively selected event, only the planet sensitivity from the portion of the light curve observed after the selection is made may be considered. This is necessary to prevent prior knowledge about the event from influencing the planet frequency measurement \citep[see][for a more detailed discussion]{Yee:2015b}.


In this work, we present the planet sensitivity analysis of two microlensing events monitored in the 2014 pilot program. The purpose is to illustrate the various issues that arise in this kind of analysis. We give an overview of our computational method in Section~\ref{sec:methods}, and present the results, i.e., the planet sensitivities of two events OGLE-2014-BLG-0939 and OGLE-2014-BLG-0124, in Section~\ref{sec:results}. Some discussion on the implications of these results is given in Section~\ref{sec:discussion}.

\begin{figure*}
\centering
\epsscale{0.8}
\plotone{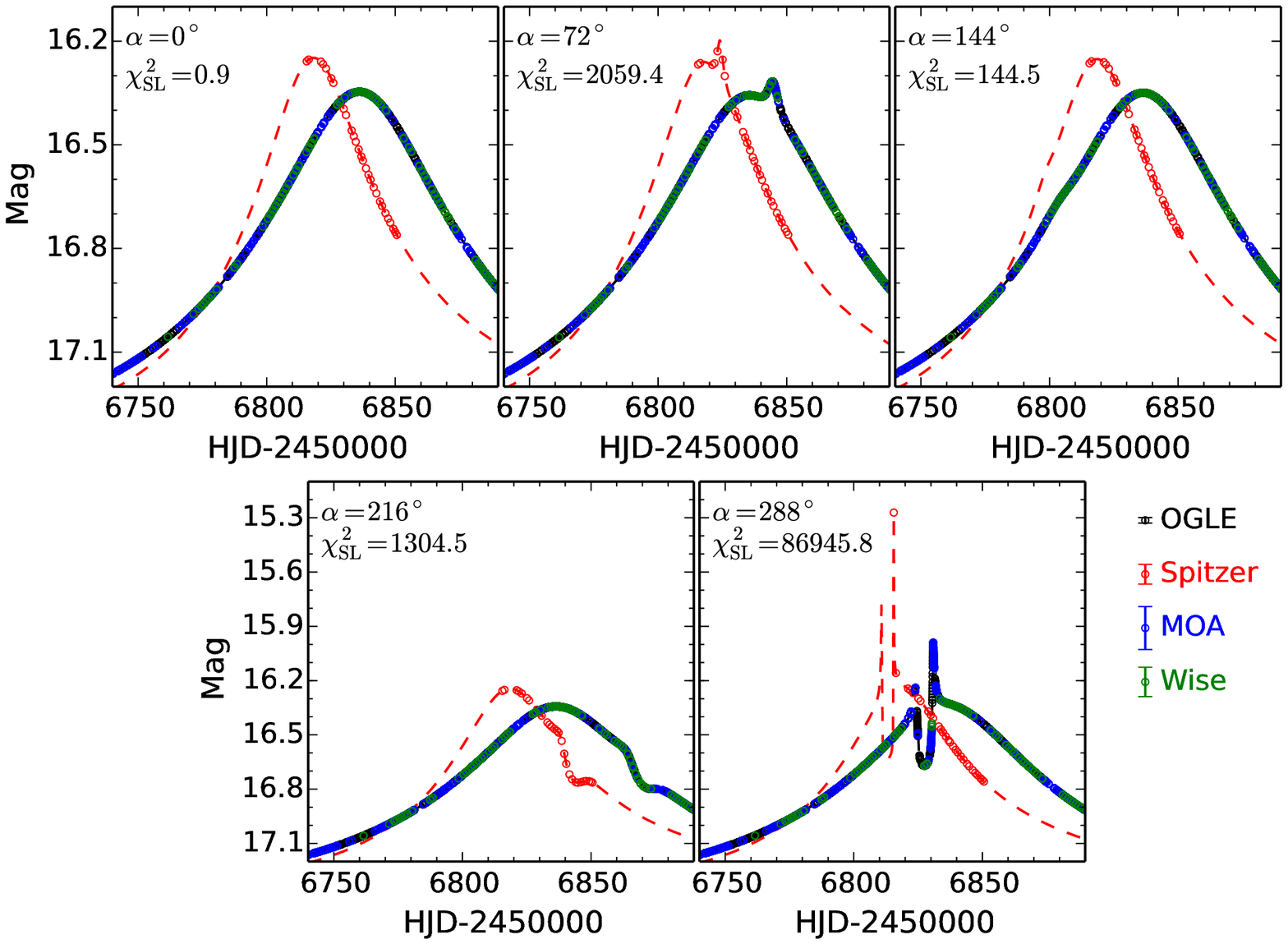}
\epsscale{0.4}
\plotone{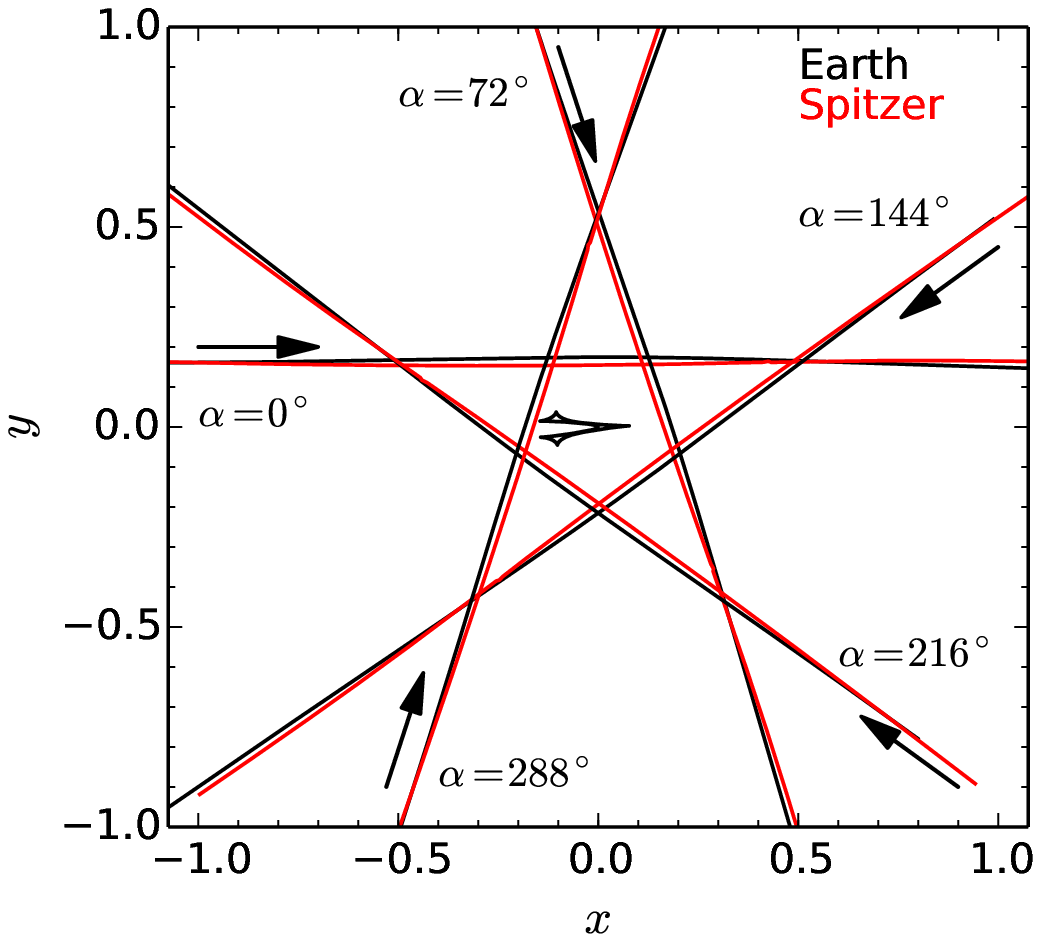}
\caption{Example light curves (upper panels) and corresponding trajectories seen from Earth and \emph{Spitzer} (lower panel) generated for the event OGLE-2014-BLG-0124. The planet-to-star mass ratio $q=7\times10^{-4}$ and the projected separation $s=0.94$ of the detected planet OGLE-2014-BLG-0124Lb are used \citep{Udalski:2015}. In the upper panel, the colors encode which team took the observation at that time, and the typical uncertainties of observations taken by different teams are indicated by the error bars to the lower right, In the lower panel, the planet and its host are placed at $(0.94,0)$ and (0,0), respectively, and the caustic arising from this lens configuration is shown in the black curve. The arrows indicate the directions of source relative motion for each $\alpha$ value. In fact, the trajectories seen from Earth and \emph{Spitzer} are fixed by the measured $\bdv{\pi}_{\rm E}$ and $u_0$ and have directions almost due east. 
While in reality, the orientation of the source trajectory is fixed on the sky and the light curves shown represent different orientations of the caustics, we choose to plot multiple sets of trajectories rather than multiple caustics in order to avoid severe overlapping among caustics.  
\label{fig:ob0124-lcs}}
\end{figure*}

\section{Methods Overview} \label{sec:methods}

The procedure for computing the planet sensitivity for microlensing events with ground-based and space-based observations has been outlined in Section~4 of \citet{Yee:2015b}. We further improve the methodology below by filling in more details about the computation.

We first measure the planet sensitivity as a function of two parameters, $S(q,s)$, where $q$ is the planet-to-star mass ratio and $s$ the projected separation normalized to the Einstein radius $\theta_{\rm E}$. The other parameters required to describe a planetary event include: the time of maximum magnification $t_0$, event impact parameter $u_0$, the Einstein timescale $t_{\rm E}$, the scaled source size $\rho$, and the planet-star axis orientation $\alpha$ (relative to the source-lens trajectory). 

We adopt the approach that was first proposed by \citet{Rhie:2000} to compute $S(q,s)$. For each set of $(q,s)$, we generate 300 planetary light curves that vary in $\alpha$ but have other parameters the same. Besides $(q,s)$, we adopt $(t_0,u_0,t_{\rm E})$ from the best fit of the single-lens/planetary event. 
We adopt the value of $\rho$ if it is measured via the finite-source effect, which is typical in planetary events (such as OGLE-2014-BLG-0124) or extremely high-magnification single-lens events. Otherwise, we choose $\rho$ as \citet{Yee:2015b} suggested. That is, combining $t_{\rm E}$, $\theta_\star$, and a reasonable choice of $\mu_{\rm rel}$ to determine $\rho$ by $\rho=\theta_\star/\theta_{\rm E}=\theta_\star/(\mu_{\rm rel}t_{\rm E})$, as in the case of OGLE-2014-BLG-0939.
\footnote{Note that while a fairly precise estimate of $\rho$ is often required to properly model real planetary microlensing events, only a rough estimate is needed to estimate the planet sensitivity of point-lens events \citep{Gaudi:2002}.}

Each planetary light curve is generated by creating fake data points at the times when the real measurements were taken, with values equal to those predicted by the model and error bars that are the same in magnitudes as those of the real data points. To maximize the efficiency of the planetary light curve computation, we use the point-source, quadrupole, and hexadecapole \citep{PejchaHeyrovsky:2009,Gould:2008} approximations when the source is approaching but still reasonably far ($\gtrsim 2 \rho$) from the caustics. For epochs that are near or on crossing caustics, we use contour integration, in which the limb-darkening effect is accommodated by using 10 annuli \citep{GouldGaucherel:1997,Dominik:1998}. This contour integration may fail under specific lens-source configurations (for example, if the lens sits on a sharp cusp, the contour integration may fail to identify some tiny images), in which case the more time-consuming inverse ray-shooting is used \citep{Dong:2006}. 
\footnote{\citet{Penny:2014} developed a new algorithm to speed up the computation of light curves with extremely low-mass planets ($q\lesssim10^{-5}$). Because of the lack of such planets in our current analysis, we end up not using this algorithm.}
We then find the best-fit single-lens model of each of these fake light curves using the downhill simplex algorithm, and quantify the deviation between the best-fit model and the fake data by $\chi^2_{\rm SL}$.
\footnote{The other approach to compute $S(q,s)$, first used in \citet{GaudiSackett:2000}, requires fitting planetary models to the real data, and thus is extremely time-consuming, although it has the advantage of simultaneously searching for all planets that may be lurking in the data down to the adopted threshold \citep{Gaudi:2002}}
In searching for the best-fit single-lens model, any unphysical model that has severely negative blending \citep[$F_{\rm B}<-0.2$, i.e., an $I<19.75$ ``anti-star'',][]{Smith:2007} is automatically rejected. Here $F=1$ is defined to correspond to $I=18$. With the absence of observational systematics and statistical fluctuations in these fake light curves, any $\chi^2_{\rm SL}$ would be due to the presence of the injected planet. We choose $\chi^2_{\rm SL}=200$ as the threshold for a ``detection'', which has been shown to be reasonable by various simulations \citep[e.g.,][]{Zhu:2014,Henderson:2014}. Then the fraction of all the $\alpha$ at fixed $(q,s)$ for which the planet is detectable is said to be the sensitivity $S(q,s)$.

In principle, one might argue that the criterion we set above for claiming a ``detection'' is only reasonable in simulated events but not enough for real planet detections, because it is possible that most of the $\chi^2$ may come from a single or two data points. This is especially true for small planets in events with low cadence, such as OGLE-2014-BLG-0939 \citep{Yee:2015a}. However, in the presence of various systematics in real microlensing observations, such an anomaly cannot be securely claimed as a detection. Thus, it may seem that other criteria are required to take this issue into account. For example, \citet{Shvartzvald:2012} requires at least three consecutive data points each with a $3\sigma$ deviation from the best-fit single-lens model to simulate the anomaly detection process in survey mode \citep[e.g.,the OGLE Early-Early Warning System anomaly detector,][]{Udalski:2003}. The problem of such a criterion is that it does not take into account the presence of potential follow-up observations. Once a small anomaly is found in the survey data, the follow-up teams will obtain more intensive observations to confirm the nature of this anomaly, but because these follow-up observations are triggered by the anomaly, they cannot be included in estimating the planet sensitivity. Therefore, we think our single criterion based on $\chi^2$ remains reasonable in a more realistic situation.

For each event, we first compute $S(q,s)$ in a grid with ten $q$ values equally spaced in $\log{q}$ from $10^{-5}$ to $10^{-2}$ and ten $s$ values equally spaced in $\log{s}$ from 0.1 to 10. Based on this preliminary $S(q,s)$ map, a finer grid is constructed and the final $S(q,s)$ map is computed. In the end, the planet sensitivity as a function of $q$ is obtained by marginalizing over all $s$ values, and the integrated planet sensitivity $S_{\rm tot}$ is given by further marginalizing over all $q$ values,
\[ S(q) = \int S(q,s) d\log{s};\ S_{\rm tot} = \int^{-2}_{-5} S(q) \xi(q) d\log{q}\ .\]
The first step assumes a flat distribution in $\log{s}$ (i.e., quasi \"Opik's law), and the second step adopts some planet mass function $\xi(q)$.

As has been emphasized in \citet{Yee:2015b}, only observations that were carried out without reference to the presence or absence of a planet can be included when $S(q,s)$ is being calculated. The treatment of this constraint varies for different events. We return to this point with individual examples in Section~\ref{sec:results}.

We apply our method to the first two events that have simultaneous observations taken from space and ground, OGLE-2014-BLG-0939 and OGLE-2014-BLG-0124.
\footnote{Although OGLE-2005-SMC-001 is the first microlensing event with a space-based microlens parallax measurement \citep{Dong:2007}, it is not included here, because its \emph{Spitzer} light curve has too sparse observations to allow a planet sensitivity analysis, and because the lens system is a stellar binary.}
For a simple demonstration of our method, we show in Figure~\ref{fig:ob0124-lcs} some fake light curves generated for the planetary event OGLE-2014-BLG-0124. The injected planet has $q=7\times10^{-4}$ and $s=0.94$, both taken as the detected values. Readers can find the real OGLE-2014-BLG-0124 event light curve in Figure~2 of \citet{Udalski:2015}.

\begin{table}
\centering
\caption{Microlensing parameters of OGLE-2014-BLG-0939 and OGLE-2014-BLG-0124 used in the planet sensitivity calculation. Parameters of OGLE-2014-BLG-0939 are taken from \citet{Yee:2015a}, and parameters of OGLE-2014-BLG-0124 are from \citet{Udalski:2015}.
\label{tab:events}}
\begin{tabular}{lccc}
\tableline\tableline
Parameters & Unit & \tabincell{c}{OGLE-2014-BLG-0939 \\ ($u_{0,-,-}$ solution)} & \tabincell{c}{OGLE-2014-BLG-0124 \\ ($u_0>0$ solution)} \\
\tableline
$t_0-6800$ & day & 36.20 & 36.176 \\
    $ u_0$ & --- & -0.913 & 0.1749 \\
    $t_{\rm E}$ & day & 22.99 & 152.1 \\
    $\pi_{\rm E,N}$ & --- & 0.220 & -0.0055 \\
    $\pi_{\rm E,E}$ & --- & 0.238 & 0.1461 \\
    $\rho$ & --- & 0.012 & 0.00125 \\
$q$ & --- & --- & 0.00694 \\
$s$ & --- & --- & 0.9443 \\
\tableline\tableline
\end{tabular}
\end{table}

\section{Results} \label{sec:results}
Below we present the planet sensitivity analysis of two events, OGLE-2014-BLG-0939 and OGLE-2014-BLG-0124, both from the 2014 \emph{Spitzer} microlens parallax program. Although follow-up observations in most cases are crucial to characterize the properties of microlensing planets, they are usually taken as responses to the anomaly in the data taken by survey teams in the case of low or moderate magnification events. Since only those observations that are taken without reference to the presence or absence of planets can be used to derive the planet sensitivity, we only include survey data in the present analysis.

The microlensing parameters of these two events, used for planet sensitivity computations, are given in Table~\ref{tab:events}.

\begin{figure*}
\centering
\epsscale{0.9}
\plotone{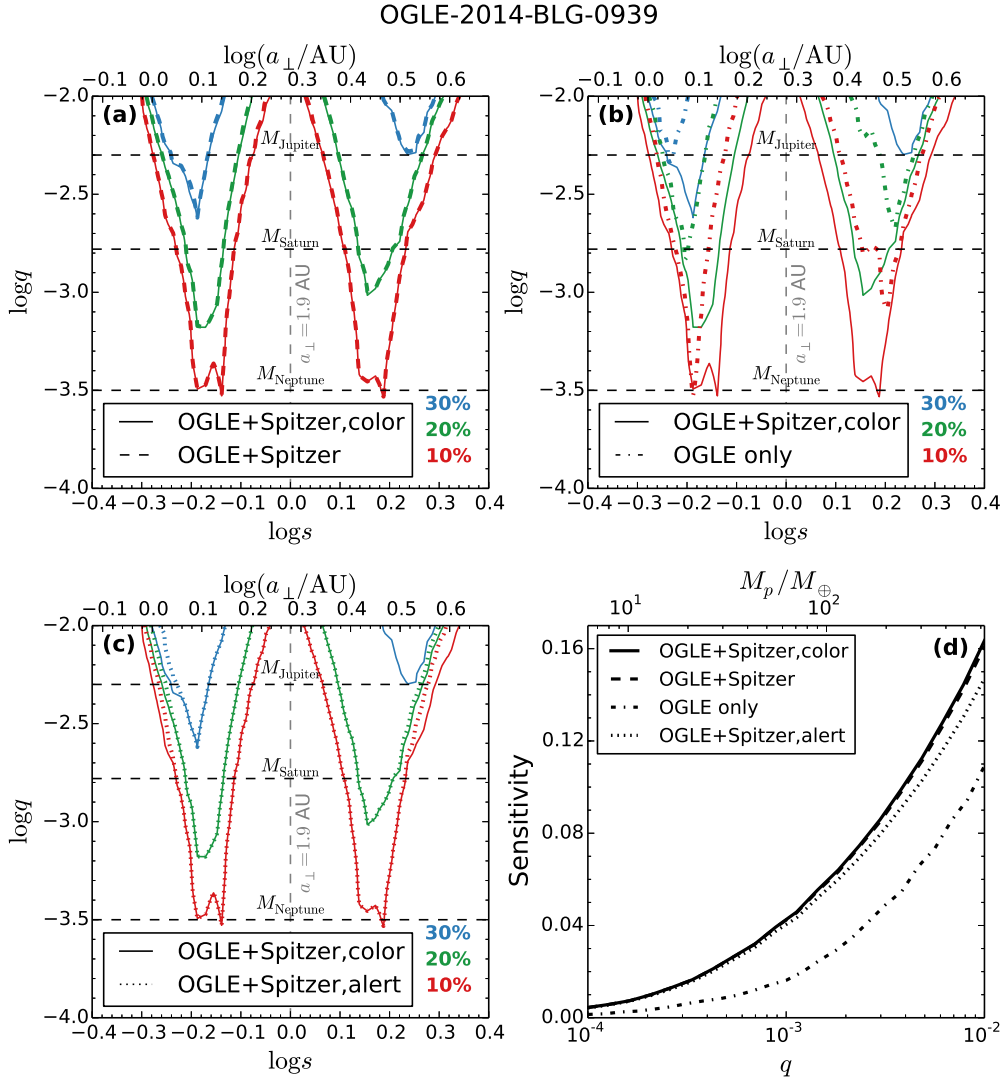}
\caption{Planet sensitivity curves of OGLE-2014-BLG-0939. The solid, dashed, dot-dashed, and dotted curves represent the sensitivity curves from ``OGLE+Spitzer,color'', ``OGLE+Spitzer', ``OGLE only'', and ``OGLE+Spitzer,alert'', respectively. In all these cases the $(-,-)$ solution, i.e., the correct solution, has been used. Panels (a), (b) and (c) show the ``OGLE+Spitzer,color'' map and one of the other three sensitivity maps, with the colors representing the curves with different sensitivities in $S(q,s)$, and the horizontal dashed lines indicating the masses of three Solar system planets. The projected separation $s=1$ corresponds to a physical separation of 1.9 AU. Panel (d) shows the four $S(q)$ sensitivity curves in a single plot.
\label{fig:ob0939-1}}
\end{figure*}

\begin{figure*}
\centering
\epsscale{1.}
\plotone{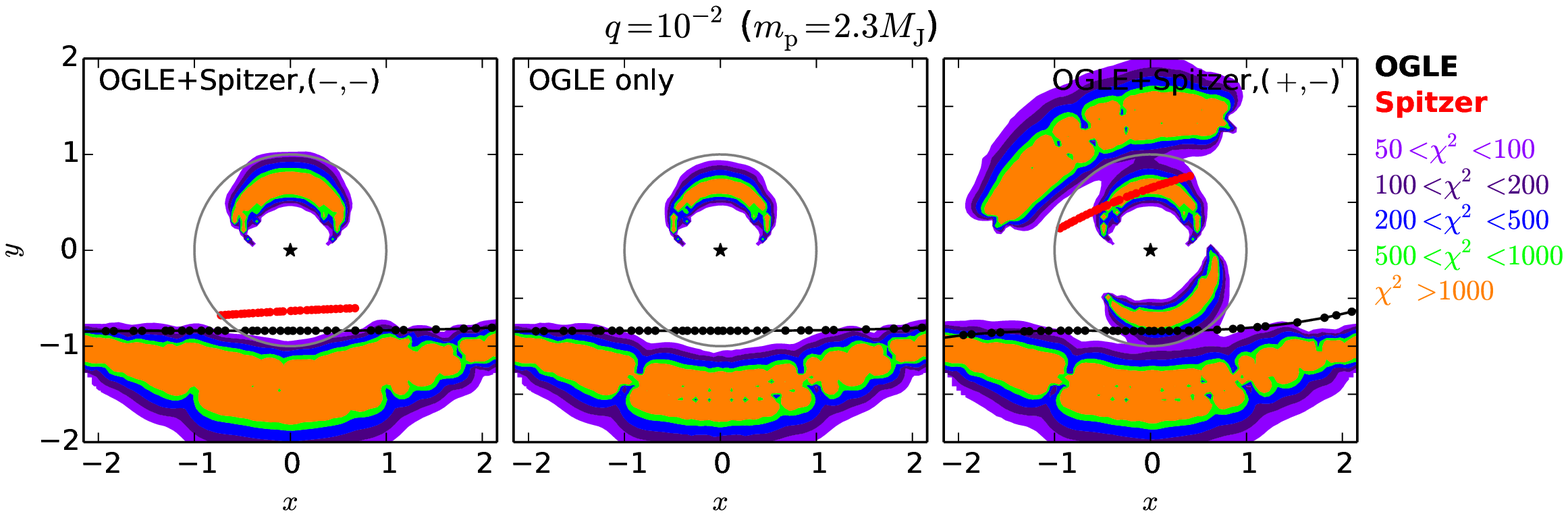}
\plotone{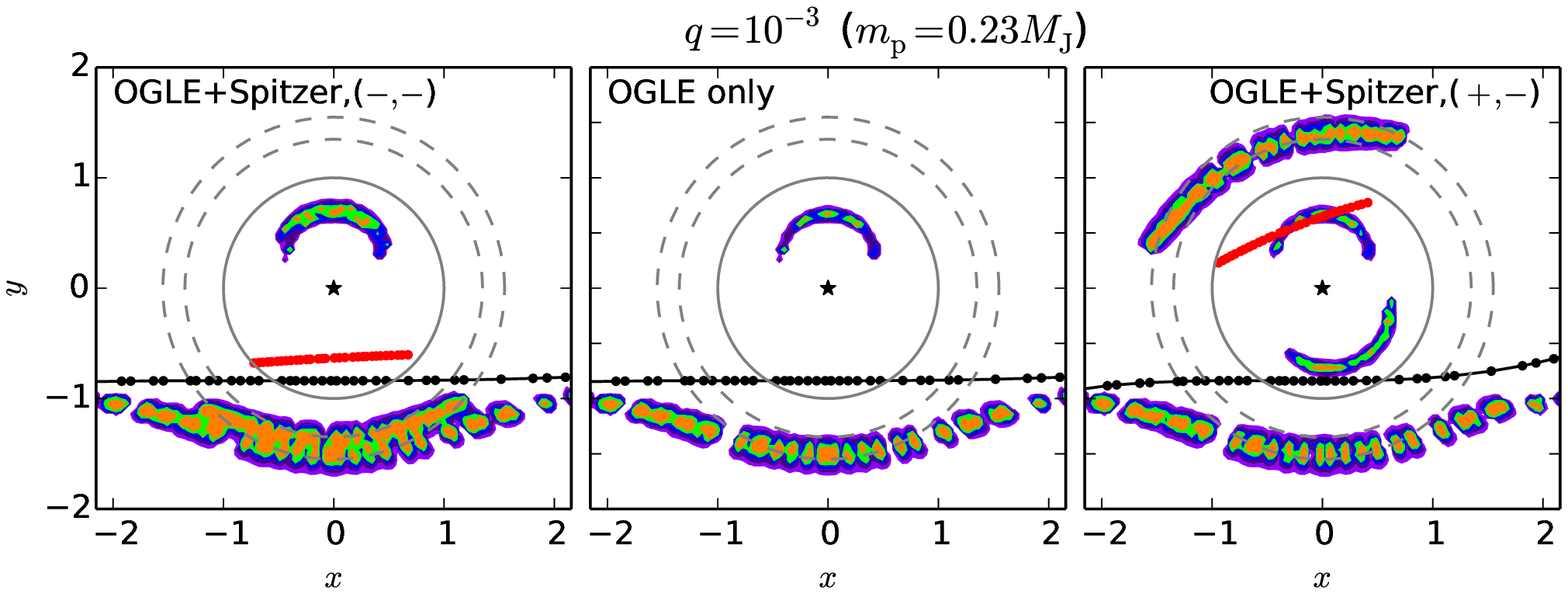}
\caption{The $\chi^2$ distributions of fake OGLE-2014-BLG-0939 light curves with planets placed at different positions $(x,y)$. The black/red lines indicate the source trajectories seen by OGLE/\emph{Spitzer}, with the dots representing the positions where the observations were taken. The lens is placed at $(0,0)$. Planets in the upper panels have mass ratio $q=10^{-2}$, and planets in the lower panels have mass ratio $q=10^{-3}$. For both mass ratios we show the $\chi^2$ distributions from ``OGLE+Spitzer,$(-,-)$'' (the same as ``OGLE+Spitzer,color'' in Figure~\ref{fig:ob0939-1}), ``OGLE only'', and ``OGLE+Spitzer,$(+,-)$ cases. The grey solid circle in each panel is the Einstein ring, and the grey dashed circles have 1.35 and 1.55 Einstein radii, which correspond to the most sensitive $s$ values for \emph{Spitzer} data and OGLE data, respectively. The discontinuities in the $\chi^2$ distribution in lower panels are caused by the discontinuity among observations. Note that the $(+,-)$ solution (rightmost panels) is included here for pedagogical purpose only and is excluded by the measurement of the lens-source relative proper motion. We denote the injected planet as ``detected'' if the fake planetary light curve has $\chi^2>200$.
\label{fig:chi2maps}}
\end{figure*}

\begin{figure*}
\centering
\epsscale{1.}
\plotone{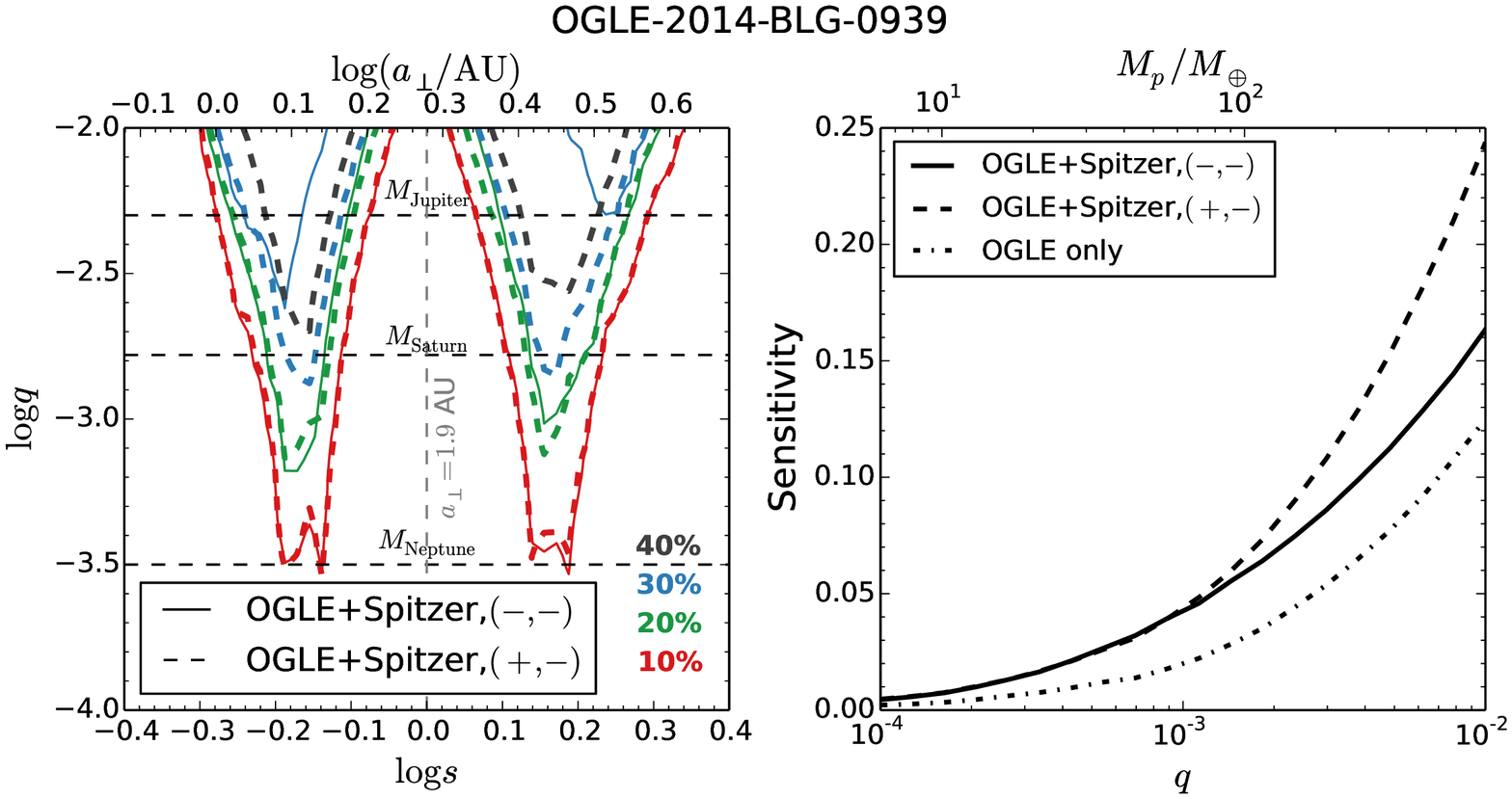}
\caption{Comparsion of planet senstivities for two degenerate geometries that might in principle be inferred from the same OGLE-2014-BLG-0939 light curve. Coding of planet sensitivity curves is similar to Figure~\ref{fig:ob0939-1}.
The solid, dashed, and dot-dashed curves represent the sensitivity curves from ``OGLE+Spitzer,$(-,-)$'' (the same as ``OGLE+Spitzer,color'' in Figure~\ref{fig:ob0939-1}), ``OGLE+Spitzer,$(+,-)$'', and ``OGLE only'' (not shown in the left panel), respectively. 
In fact, the (+,-) solution was ruled out by \citet{Yee:2015a}, but is shown here to illustrate the impact of this type of degeneracy.
\label{fig:ob0939-2}}
\end{figure*}

\subsection{Single-lens event OGLE-2014-BLG-0939}
OGLE-2014-BLG-0939 was the first single-lens event with space-based microlens parallax measurement \citep{Yee:2015a}. The four-fold degeneracy, arising from the fact that 
the projected position of the source onto the sky as seen from Earth and the satellite can pass by the lens from the same side or different sides
\citep{Refsdal:1966,Gould:1994}, was effectively broken given the measurement of the source proper motion, yielding a star with mass $M=0.23M_\odot$ at distance $D_{\rm L}=3.1$ kpc. However, since the four-fold degeneracy is a generic feature of most single-lens events, we also consider below the case in which it would not have been broken.

The event OGLE-2014-BLG-0939 was only observed by the Optical Gravitational Lensing Experiment \citep[OGLE,][]{Udalski:2003}. It lies in OGLE field BLG630, implying that it was observed at relatively low cadence, roughly once per two nights. We include OGLE data taken from HJD$'\equiv$ HJD $- 2450000=6000$ until the end of the 2014 season (HJD$'$ $\approx$ 6941). These boundaries are somewhat arbitrary, but allow for a substantial baseline while keeping the number of data points manageable. After the removal of isolated outliers, we find in total 248 observations. This event also received in total 31 \emph{Spitzer} observations during the interval HJD$'$ = 6814.1 to 6845.7, with close to once per day cadence. Readers can find more details about these observations in \citet{Yee:2015a}.

Although event OGLE-2014-BLG-0939 was not expected to show substantial planet sensitivity because of the low observational cadence, it is chosen for the present analysis for other reasons. First, as the first analysis of planet sensitivity from combined ground- and space-based microlensing observations, it is natural to use the first single-lens event that allows us to do so. Second, \emph{Spitzer} captured the peak of this event with a cadence higher than that of OGLE, and has an impact parameter ($u_0=0.6$) smaller than that of OGLE ($u_0=0.9$), so even though \emph{Spitzer} has fewer observations overall, they are expected to contribute a significant fraction to the overall planet sensitivity. This helps us partly in testing the numerical result with our intuition based on microlensing theory, but mostly in understanding some related issues, as we will see below. Finally, the planet sensitivity of the first single-lens event with space-based parallax measurement has historical interest.

We first construct four different planet sensitivity maps $S(q,s)$ of OGLE-2014-BLG-0939 by treating this event under various different assumptions. First, we find $S(q,s)$ with parameters listed in Table~\ref{tab:events}, using both OGLE and \emph{Spitzer} data, with no constraint; this sensitivity map is marked as ``OGLE+Spitzer'' in Figure~\ref{fig:ob0939-1}. As \citet{CalchiNovati:2015} has pointed out, one can find the constraint on the source flux in \emph{Spitzer} 3.6 $\mu$m band by doing a linear regression between $V-I$ and $I-[3.6{\ \rm \mu m}]$ colors for non-microlensed field stars, and this flux constraint helped in a few cases to break the four-fold degeneracy. Therefore, we also compute $S(q,s)$ with this flux constraint imposed, with the result marked as ``OGLE+Spitzer,color'' in Figure~\ref{fig:ob0939-1}, and consider this $S(q,s)$ as the final planet sensitivity map of this event and the standard against which the others should be compared.
In the third test, we use only OGLE data, and the $S(q,s)$ based on this is called ``OGLE only''. 

In the above cases we treat this event as selected for \emph{Spitzer} observations objectively. However, as has been shown in \citet{Yee:2015b}, events can also be selected for space-based parallax measurements subjectively. This is the case for OGLE-2014-BLG-0939, because there was no pre-determined objective criteria for 2014 season events. Therefore, we also consider the case in which this event was selected subjectively.
We set the date when it was chosen for \emph{Spitzer} observations to be the public announcement date, or the alert date \citep{Yee:2015b}. For OGLE-2015-BLG-0939, $t_{\rm alert}=6811$. We then compute $S(q,s)$ following the procedure proposed by \citet{Yee:2015b}, which we described in Section~\ref{sec:methods}. For each $\alpha$ at the chosen $(q,s)$, we first fit the fake data that were released before $t_{\rm alert}$ and find $\chi^2_{\rm SL,alert}$. If $\chi^2_{\rm SL,alert}>10$, we regard the injected planet as having been noticeable, and thus reject this $\alpha$; otherwise we fit the whole data set to find $\chi^2_{\rm SL}$ and compare it with the threshold we choose to determine the detectability at this $\alpha$. The sensitivity map computed in this way is marked as ``OGLE+Spitzer,alert''.

The most prominent feature in Figure~\ref{fig:ob0939-1} is the two separated triangular structures that are nearly symmetric about $s=1$, which is a feature of low-magnification events \citep[see, e.g., Figure~8 of ][]{Gaudi:2002}. The second remarkable feature in all but the ``OGLE only'' $S(q,s)$ maps is the double-peak structure at the bottom of each triangle diagram. This is better illustrated in Figure~\ref{fig:chi2maps}, in which we present the spatial distributions of $\chi^2$ at various planet positions for two chosen $q$ values. The reason for this feature is that the event was seen from two observatories to have fairly different impact parameters (0.6 vs. 0.9) leading to two values of $s$ at which the planet sensitivity peaks. For low and intermediate magnification events, the value of $s$ at which the $S(q,s)$ map peaks is set by \citep{Gaudi:2002}
\[ \left| s-\frac{1}{s} \right| = u_0\ .\]
This gives us $\log{s} = \pm 0.13$ for \emph{Spitzer}, and $\log{s} = \pm 0.19$ for OGLE, both consistent with our numerical result.

As seen in panel (b) of Figure~\ref{fig:ob0939-1}, the planet sensitivity changes dramatically from excluding to including the \emph{Spitzer} light curve, as has been expected. We find no noticeable improvement in planet sensitivity by imposing the color constraint, and no significant loss of sensitivity when treating this event as subjectively selected. The reason for both is that the \emph{Spitzer} light curve captured the peak of the event: \citet{CalchiNovati:2015} showed that only in events in which \emph{Spitzer} did not capture the peak could the color information improve the single-lens fit. The same applies to the planet sensitivity. Because the peak contributes most of the planet sensitivity, excluding data before $t_{\rm alert}<t_{0,Spitzer}$ can only marginally modify the sensitivity curves by reducing sensitivity to planets farther from the lensing zone ($0.6-1.6 R_{\rm E}$).

The four-fold degeneracy is a generic feature of single-lens events if observed by two well-separated observers \citep{Refsdal:1966,Gould:1994}. Although it was broken in the present case by the measurement of the source proper motion, we consider here what the planet sensitivity would be if this four-fold degeneracy had persisted. Since this four-fold degeneracy reduces to a two-fold degeneracy when one ignores the direction of motion and only consider the mass and distance of the lens system, one only need consider two of the four solutions. In the case of OGLE-2014-BLG-0939, these are the ($-,-$) solution, i.e., the correct solution, and its counterpart ($+,-$) solution.
\footnote{A better way to mark the four solutions was introduced in \citet{Zhu:2015a}. However, this old label system is still used here in order to keep the consistency with \citet{Yee:2015a}. The four solutions $(-,+)$, $(+,+)$, $(-,-)$, and $(+,-)$ by the definition of \citet{Yee:2015a} and \citet{CalchiNovati:2015} correspond to $(+,+)$, $(+,-)$, $(-,-)$, and $(-,+)$ solutions by the definition of \citet{Zhu:2015a}, respectively.}
The $(-,-)$ solution has Earth and \emph{Spitzer} trajectories pass from the same side of the lens, whereas the $(+,-)$ solution has the two trajectories pass from different sides of the lens. See Figure~\ref{fig:chi2maps} for an illustration of the source trajectories seen from Earth and \emph{Spitzer} for these two different solutions. Figure~\ref{fig:ob0939-2} shows the sensitivity map from this $(+,-)$ solution in comparison with that from the $(-,-)$ solution (i.e., ``OGLE+Spitzer,color'' in Figure~\ref{fig:ob0939-1}) and that of ``OGLE only''. It shows that the planet sensitivity from the $(+,-)$ solution is considerably higher for giant planets ($q>10^{-3}$) with respect to the $(-,-)$ solution, and almost doubles that from ``OGLE only''. The reason is that the OGLE and \emph{Spitzer} light curves are sensitive to different $\alpha$ values in the $(+,-)$ solution. In cases for which the degeneracy cannot be resolved, we recommend taking the mean of the sensitivities from these two solutions, weighted by the likelihood for each solution to be the correct one, for example the $\chi^2$ difference and the Rich argument \citep{CalchiNovati:2015}.

\subsection{Planetary event OGLE-2014-BLG-0124} \label{sec:ob0124}

\begin{figure*}
\centering
\epsscale{1.}
\plotone{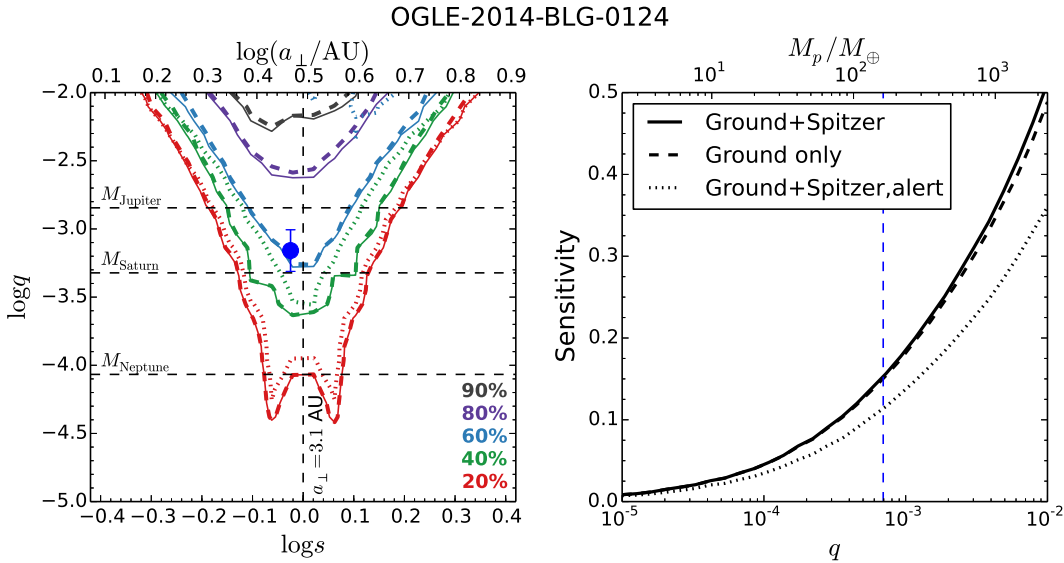}
\caption{Planet sensitivity curves of OGLE-2014-BLG-0124. The left panel shows the $S(q,s)$ map, and the right panel shows the $S(q)$ distribution. The solid, dashed, and dotted curves represent the sensitivity curves from ``OGLE+Spitzer'', ``OGLE only'', and ``OGLE+Spitzer,alert'', respectively. The position of the detected planet OGLE-2014-BLG-0124Lb is indicated as a blue dot with error bar in the left panel and a vertical blue dashed line in the right panel. The projected separation $s=1$ in this case corresponds to a physical separation of 3.1 AU.
\label{fig:ob0124}}
\end{figure*}

OGLE-2014-BLG-0124 is the first microlensing planetary system with a space-based parallax measurement \citep{Udalski:2015}. About 4 kpc away from Earth, the planet has mass $0.5 M_{\rm J}$, and is separated from a $0.7 M_\odot$ star by $a_\perp\approx 3.1$ AU.

The event OGLE-2014-BLG-0124 lies in OGLE field BLG512, meaning that it was observed at OGLE's highest cadence, about once every 20 minutes. Again, we only include data that were taken after HJD$'$ = 6000 but before the end of the 2014 season, and find in total 6647 OGLE observations. In addition, this event also falls into the microlensing fields of the Microlensing Observations in Astrophysics \citep[MOA,][]{Sako:2008} and the Wise Observatory \citep{Shvartzvald:2012} surveys, and therefore received 8865 and 1010 observations from MOA and Wise respectively during that time interval. Since this is a planetary event, intensive follow-up observations were taken after the planet anomaly was noticed.
\footnote{Among all follow-up teams, only the Microlensing Follow-Up Network \citep[$\mu$FUN,][]{Gould:2010} obtained a few observations before this event became anomalous, but the purpose of these a few observations was to better characterize the source rather than to find planets.}
However, as has been emphasized before, these follow-up observations cannot be used for the planet sensitivity analysis. We mark all the ground-based observations from the three survey groups as the ``ground'' data set.

OGLE-2014-BLG-0124 received in total 39 \emph{Spitzer} observations during the interval HJD$'$ = 6815.5 to 6850.6. See Figure~1 of \citet{Udalski:2015} for the distribution of these observations. Although these \emph{Spitzer} observations play a significant role in measuring the parallax effect, as \citet{Udalski:2015} has shown, they are expected to contribute only a tiny fraction to the overall planet sensitivity because the trajectory of the source relative to the lens is very similar as seen from the ground and from \emph{Spitzer}. Even so, however, we cannot include all 39 \emph{Spitzer} observations for the planet sensitivity analysis, because some of them were carried out in response to the planetary anomaly found in the ground-based data. Considering that the \emph{Spitzer} cadence was about once per day before the planet anomaly alert, we remove some \emph{Spitzer} observations to form a time sequence with similar cadence. In the end, 29 \emph{Spitzer} observations are included. Some example fake light curves generated from these two data sets are shown in Figure~\ref{fig:ob0124-lcs}.

We construct three sensitivity maps for OGLE-2014-BLG-0124: one with only ground-based data, one with both ground and \emph{Spitzer} data but no color constraint, and one by treating it as subjectively selected. We do not consider the impact of the color constraint, since the previous case has shown that if the peak of the event seen from \emph{Spitzer} was captured, the color constraint can only modify the sensitivity curve marginally. We do not consider the case of four-fold degenerate solutions here either, because the opposite solution, which would have $\pi_{\rm E,N}\sim0.35$, could be effectively ruled out by the ground-based data alone even if the planet were not detected. As with OGLE-2014-BLG-0939, in the case in which we treat the event as subjectively chosen, the date when this event was chosen for \emph{Spitzer} observation is used as the alert date, i.e., $t_{\rm alert}=6811$.

Figure~\ref{fig:ob0124} shows these sensitivity maps $S(q,s)$ and marginalized sensitivities $S(q)$, with the position of the detected planet OGLE-2014-BLG-0124Lb ($q=7\times10^{-3}$ and $s=0.94$) marked for reference. As expected, although \emph{Spitzer} observations are key to measuring the parallax effect, the removal of the entire \emph{Spitzer} light curve has negligible effect on the overall planet sensitivity. In the case that this event was selected subjectively, the sensitivity drops by up to $\sim 30\%$, but still remains significant, because the alert date $t_{\rm alert}$ was fairly far from the event peaks as seen from Earth.

\section{Discussion} \label{sec:discussion}

We present the first planet sensitivity analysis for microlensing events with simultaneous observations taken from space and ground. We apply our methodology to two such events, the single-lens event OGLE-2014-BLG-0939 \citep{Yee:2015a} and the planetary event OGLE-2014-BLG-0124 \citep{Udalski:2015}. 
Assuming an underlying planet population of one planet per dex$^2$ per star,
we find that OGLE-2014-BLG-0939 with a lens distance of 3.1 kpc shows an overall 11.5\% detection efficiency for planets with mass ratio in the range $10^{-4}<q<10^{-2}$ (or mass range from $6M_\oplus$ to $2M_{\rm J}$), and that OGLE-2014-BLG-0124 with a lens distance of 4 kpc has 45\% detection efficiency for planets with $10^{-5}<q<10^{-2}$ (or mass range from $2M_\oplus$ to $7M_{\rm J}$). 

The contributions to the overall planet sensitivity from space-based observations are considerably different in the two events. For OGLE-2014-BLG-0939, although \emph{Spitzer} took fewer observations than OGLE did, the space and ground observations contribute nearly equally to the overall sensitivity, because the event had higher magnification as seen from \emph{Spitzer} and was observed by \emph{Spitzer} with a higher cadence. This result means that the planet sensitivity could have been higher if the solution with the source passing by the lens from different sides as seen from \emph{Spitzer} and Earth were allowed, because then \emph{Spitzer} would probe a significantly different region surrounding the lens (see the rightmost panels of Figure~\ref{fig:chi2maps}). It also suggests that a dedicated microlensing survey from space, when combined with ground-based surveys, can dramatically increase the probability to find planets, besides its capability to efficiently measure the microlens parallax effect \citep{GouldHorne:2013}. For OGLE-2014-BLG-0124, the ground-based observations are much more intensive and cover a much longer baseline, leading to a negligible contribution to the planet sensitivity from \emph{Spitzer} data, although these \emph{Spitzer} observations are key to measuring the microlens parallax.

Although neither of these two events showed high-magnification behavior, they both showed substantial planet sensitivities as seen in Figures~\ref{fig:ob0939-1} and \ref{fig:ob0124}. This result has implications for conducting microlensing experiments using narrow-angle space-based satellites such as \emph{Spitzer}. Limited by their field-of-view, such narrow-angle satellites can only be used to follow up events that were found from the ground. Ideally, one would want to follow up events that are very sensitive to planets, for example, high-magnification ($A_{\rm max}>100$) events. However, two intrinsic difficulties of high-magnification events render such a strategy difficult to implement. First, high-magnification events are intrinsically very rare: one will only expect to have a few within the $\sim38$ day Bulge window for the \emph{Spitzer} telescope. Second, due to operational constraints, using \emph{Spitzer}-like telescopes for microlens parallax measurements requires one to choose events at least a few days before the observations start.
\footnote{It is about $3-9$ days for \emph{Spitzer}. See Figure~1 of \citet{Udalski:2015} for an illustration of the observing strategy with \emph{Spitzer}.}
However, it is difficult to confidently identify events that will achieve high magnification in advance. Our result here suggests that an ensemble of the more common low to moderate magnification events can also yield significant planet sensitivity and therefore probability to detect planets.

Another key component of doing microlensing experiments using telescopes such as \emph{Spitzer} is event selection. To make a well controlled sample of events suitable for statistical studies, it is best to select all events based on some objective criteria. However, subjectively choosing events is also necessary for two good reasons. First, the objective criteria cannot capture all events of interest, and second, an earlier subjective trigger may make the difference between measuring a parallax or not \citep{Yee:2015b}. As illustrated in Figure~\ref{fig:ob0939-1}c, the planet sensitivity of OGLE-2014-BLG-0939 considering only the data after the event is subjectively selected is comparable to the full planet sensitivity. In this case, the reason is that a significant fraction of the planet sensitivity comes from the Spitzer data themselves. However, in general, this illustrates the value of subjective event selections even though only part of the light curve may be considered when calculating planet sensitivity.

The analysis of these two events illustrates the role of space-based observations of microlensing events in measuring planet sensitivity, which is one component of measuring the Galactic distribution of planets. First, the additional observations from well-separated observatory probe a different region of the lens system, increasing the planet sensitivity. Second, this demonstrates that low and moderate magnification events show substantial planet sensitivity. Both of these points apply to space-based microlensing parallax observations in general including the K2 microlensing campaign. Finally, we have shown that even though events may be selected subjectively for a targeted campaign like \emph{Spitzer}'s, they may still contribute significant planet sensitivity to the final sample as suggested by \citet{Yee:2015b}.

\acknowledgments
Work by WZ and AG was supported by JPL grant 1500811.
Work by JCY and MP was performed under contract with the California Institute of Technology (Caltech)/Jet Propulsion Laboratory (JPL) funded by NASA through the Sagan Fellowship Program executed by the NASA Exoplanet Science Institute.
Work by Y.S. is supported by an appointment to the NASA Postdoctoral Program at the Jet Propulsion Laboratory, administered by Oak Ridge Associated Universities through a contract with NASA.
The OGLE project has received funding from the National Science Centre, Poland, grant MAESTRO 2014/14/A/ST9/00121 to AU.
TS acknowledges the financial support from the JSPS23103002, JSPS24253004 and JSPS26247023. The MOA project is supported by the grant JSPS25103508 and 23340064.
This work is based in part on observations made with the Spitzer Space Telescope, which is operated by the Jet Propulsion Laboratory, California Institute of Technology under a contract with NASA.

\end{CJK*}
\end{document}